\documentclass[a4paper,reqno,superscriptaddress,12pt]{revtex4}

\usepackage{graphicx}
\usepackage{dcolumn}
\usepackage{epsfig}

\usepackage[centertags]{amsmath}
\usepackage{amsfonts}
\usepackage{euscript}
\usepackage{amssymb}
\usepackage{amsthm}
\usepackage{newlfont}
\usepackage{stmaryrd}
\usepackage{mathrsfs}
\usepackage{subfigure}

\let\al=\alpha \let\be=\beta \let\de=\delta \let\ep=\epsilon
\let\ve=\varepsilon   
 \let\la=\lambda  
\let\si=\sigma   
\let\ze=\zeta 

  \let\La=\Lambda \let\Om=\Omega

\newcommand{\caE}{{\mathcal E}}

\newcommand{\opunit}{\text{1}\kern-0.22em\text{l}}

\DeclareMathAlphabet{\mathpzc}{OT1}{pzc}{m}{it}

\newcommand{\id}{\textrm{d}}

\begin{document}

\title{A nonequilibrium extension of the Clausius heat theorem}
\author{Christian Maes}
\email{christian.maes@fys.kuleuven.be}
\affiliation{Instituut voor Theoretische Fysica, KU Leuven, Belgium}
\author{Karel Neto\v{c}n\'{y}}
\email{netocny@fzu.cz}
\affiliation{Institute of Physics,  Academy of Sciences of the Czech Republic, 18221 Prague, Czech Republic}

\begin{abstract}
We generalize the Clausius (in)equality to overdamped mesoscopic and macroscopic diffusions in the presence of nonconservative forces. In contrast to previous frameworks, we use a decomposition scheme for heat which is based on an exact variant of the Minimum Entropy Production Principle as obtained from dynamical fluctuation theory. This new extended heat theorem holds true for arbitrary driving and does not require assumptions of local or close to equilibrium. The argument remains exactly intact for diffusing fields where the fields correspond to  macroscopic profiles of interacting particles under hydrodynamic fluctuations.
We also show that the change of Shannon entropy is related to the antisymmetric part under a modified time-reversal of the time-integrated entropy flux.
\end{abstract}


\maketitle

\begin{center}
{\large {\bf in honor of Herbert Spohn on the occasion of his 65th birthday }}
\end{center}

% =================================================================================
\section{Introduction}

The Clausius heat theorem stands at the very beginning of the theory of equilibrium thermodynamics~\cite{cla}.  Grown out of meditations
on the efficiency of heat machines~\cite{car}, Clausius discovered that for quasi-static processes between equilibria, heat divided by temperature is an exact differential, memorized as $\delta Q/T = \id S$. The new state function $S$ was called the entropy of the system (in equilibrium). Clausius continued by arguing that this entropy never decreases for systems in thermal isolation, or, more generally for arbitrary processes in which the system moves from an initial to a final equilibrium  holds
\[
  S_\text{fin} - S_\text{ini}
  \geq \int_{\text{ini}}^{\text{fin}} \frac{\delta Q}{T}
\]
where $\delta Q$ is the incoming heat while in consecutive contact with thermal baths at temperature $T$
(and again, with equality for quasi-static processes). From the point of view of statistical mechanics and in particular in retrospect from the equilibrium fluctuation theory developed by Boltzmann, Planck and Einstein, that Clausius relation appears challenging.  Indeed, thinking of entropy as the logarithmic volume of regions on the constant energy surface corresponding to some macroscopic condition, inscribed as $S = k_B\,\log W$, the relation with heat is less evident.  The fact that basically the same entropy covers quite different physical aspects for processes satisfying detailed balance (including also its role in defining thermodynamic forces and in the fluctuation-dissipation theorem) does of course not imply that the Clausius entropy and the Boltzmann entropy, even in their versions for open systems, must remain strictly connected also away from equilibrium when currents of particles or energy are maintained, \cite{statu}.
Yet it was found by Komatsu, Nakagawa, Sasa and Tasaki that close-to-equilibrium, in fact for Markov processes satisfying local detailed balance for which the driving is small, a slightly modified Clausius relation is valid \cite{knst} --- the basic reason there is the approximate validity of the MacLennan-Zubarev ensemble for which the occupations are still determined by heat \cite{KN} or, as we will show, the approximate validity of the Minimum Entropy Production Principle (MEPP)~\cite{min}.
Still another proposed generalization, even formally exact arbitrarily far from equilibrium though offering less direct interpretation and applications, was earlier given by Hatano and Sasa~\cite{hs}, by using an intricate decomposition of heat into steady-state and excess components.

The present paper proposes
a new
heat renormalization under which the Clausius relation continues to hold outside the linear regime around equilibrium, and
which in a sense enjoys virtues of the two proposals mentioned before, at least for diffusions in the overdamped regime. Starting by the observation that the limited validity of the generalized Clausius equality in~\cite{knst} is intimately related to the breaking of the MEPP, we present a modification which remains exactly valid. Such a modification  is indeed
possible
by employing dynamical fluctuation theory, which provides a general framework under which variational functionals are constructed as large-fluctuation potentials. Though presently restricted to overdamped diffusions where a simple interpretation emerges, this new approach offers a more systematic way of generalizing various equilibrium-like concepts and relations which goes beyond  trial-and-error. It also shows the relevance of large fluctuations for the energetics of mesoscopic thermodynamic machines.\\

The plan of the paper is then as follows.  We start in the next section with the meaning of extending Clausius heat theorem to nonequilibrium; we discuss thermodynamic transformations and we introduce in general terms the core of the argument.  A review of previous results on
mesoscopic nonequilibrium extensions of
Clausius' heat theorem is presented
in Section~\ref{prev}.
These are mostly found in the project of steady state thermodynamics carried forward in~\cite{knst,KN,hs,house}.  We restrict ourselves also there to the set-up of overdamped diffusions.
The basic inspiration of the present paper is to pass via dynamical fluctuation theory, in particular dealing with the fluctuations of the occupation times. That is explained in Section \ref{modc} and in \ref{dync} in particular.
An extension to macroscopic boundary-driven diffusion in a form which is similar yet fundamentally different from the recent contributions~\cite{roma12,roma13} will be given in Section~\ref{mab}.  The last Section \ref{trco} takes another point of view and delivers an exact Clausius equality obtained by the appropriate time-reversal on the entropy flux.\\

Herbert Spohn has certainly been a pioneer of bringing dynamical large deviation theory to shed light on nonequilibrium questions.  One of the many examples are found in Section 3.7 in his book \cite{lar} where he finds the entropy production as a path-wise quantity (dynamical contribution to the rate function) in the nonequilibrium action.  That is actually crucial in the fluctuation-understanding of the minimum entropy production principle, as used also in the present paper.

\section{The meaning of extending Clausius heat theorem}

As is often the case, extending an equilibrium idea or an equilibrium relation to nonequilibrium is not uniquely defined and its relevance is not immediate.  The Clausius heat theorem can be read in different ways for the purpose of extensions.  On the one hand, one remains interested in the question what kind of excess heat over temperature gives rise to an exact differential, and in the general question of thermal properties of nonequilibrium thermodynamic processes.  That is asking for extensions to nonequilibrium calorimetry and to discover thermodynamic potentials that generate thermal response and conduction also in nonequilibrium systems, \cite{heat}.  On the other hand, one is interested in the possible thermodynamic relevance and interpretation of
the Shannon or other entropic measures of the nonequilibrium statistics for stochastic systems, as well as of the local equilibrium (Gibbsian) entropies on the macroscopic scale.
The more general challenge is here to get operational thermodynamic meaning to the static fluctuation functionals (sometimes called nonequilibrium free energies) for nonequilibrium systems.

\subsection{Thermodynamic transformations}\label{ttr}
Central to the Clausius heat theorem is the notion of thermodynamic transformation.  One considers a number of control parameters $\alpha$ such as the temperature of the environment and the volume of the system and one watches the system as these are being changed.  In the original version, that is between an initial and a final equilibrium condition. When the transformation is quasi-static, then at each moment the condition of the system remains that of instantaneous equilibrium. In equilibrium statistical terms the Clausius equality then reads
\begin{equation}\label{gi}
    \beta\Bigl( \id \langle E_v\rangle -
    \Bigl\langle \frac{\partial E_v}{\partial v} \Bigr\rangle \id v \Bigr)
    = \id(\log Z(\alpha) + \beta \langle E_v \rangle),\quad
    \alpha=(v,\beta)
\end{equation}
 connecting canonical equilibria $\rho_\alpha(x) = \exp[-\beta E_v(x)]/Z(\alpha)$ at inverse temperature $\beta$ for energy function $E_v$ for example depending on a length scale $v$. The left-hand side is heat over temperature $\beta \delta Q$, as follows from the first law of thermodynamics.
The right-hand side is the change $\id S$ in (physical) entropy of the equilibrium system as given by the Shannon entropy of the Gibbs canonical distribution.
For more general transformations (not quasi-static)
the usual inequality $\beta \delta Q \leq \id S$ applies in \eqref{gi}.\\
Exactly because its central subject is to connect various equilibria, the Clausius statement naturally relates with questions of relaxation to and response in equilibrium.  Nature appears to accompany these processes of relaxation and response with dissipation, as also appears in the fluctuation--dissipation theorem.

Nonequilibrium thermodynamics wants to consider processes that connect stationary nonequilibria, where besides the more standard control in terms of environment temperature(s) or volume, now also manipulations of the driving condition can be part of the scheme.  Again aspects of relaxation to and thermal response in nonequilibrium stationary regimes will be crucial although the latter is far from obvious in our present approach.
A first difference here with equilibrium thermodynamics is the presence of heat also under stationary nonequilibrium. Even when the system runs stationary, a flow of energy or particles is maintained, giving rise to Joule heating or dissipation. That will lead to questions of renormalization schemes that define excess heat so to pick up exactly and only that heat which is (entirely) due to the transforming between nonequilibria.  There are however good reasons to think that heat and dissipation as formulated in irreversible thermodynamics will
no longer be sufficient
to characterize these relaxation processes. That is the more important second difference with equilibrium.  We believe that kinetic aspects such as those summarized under the name of dynamical activity will become important, and indeed some of that is visible in linear response around nonequilibria, \cite{zrp}.  That will also be taken up in Section \ref{dync}.  Yet, while our approach stems from considerations on the dynamical activity, for the purpose of staying close to the original meaning of the Clausius theorem we restrict ourselves here exactly to a set-up where the dynamical activity (excess) is precisely given by the entropy production (excess).\\

On the more technical side of the question is the mathematical modeling of thermodynamic transformations. That is monitored via the control parameter which makes the time-evolution of the dynamical variables time-inhomogeneous.

The dynamical variables are reduced variables such as macroscopic observables, profiles or the mechanical states of subsystems. To give an example in the context of Markov processes, we consider the Master equation for probability distributions $\mu_t$,
\begin{equation}\label{mas}
\dot{\mu}_t = L_{\alpha(t)} \mu_t, \qquad
0 \leq t \leq \tau
\end{equation}
where the linear operator
$L_\alpha$ is the forward generator of a Markov process for each fixed value of $\alpha$.
We assume a smooth dependence and evolution with a unique stationary distribution $\rho_\alpha$ satisfying $L_\alpha \rho_\alpha=0$ for each $\alpha$.\\
To incorporate quasi-static changes, we take
\begin{equation}\label{qst}
\alpha^\varepsilon(t) := \alpha(\varepsilon t), \qquad
0 < \varepsilon \ll 1,\,\,
%t\in [0,\tau/\varepsilon]
0 \leq t \leq \varepsilon^{-1}\tau
\end{equation}
and denote the solution of the Master equation \eqref{mas}
by
$\mu^{(\varepsilon)}_t$ for $\mu^{(\varepsilon)}_{t=0}  = \rho_{\alpha(0)} = \rho_{\text{ini}}$,
where we start from a stationary distribution for an initial choice of control parameter.  Under the quasi-static limit $\varepsilon\downarrow 0$ the probabilities $\mu^{(\varepsilon)}_t$
tend to the stationary nonequilibria, with correction
\begin{equation}\label{qsc}
\mu^{(\varepsilon)}_{\varepsilon^{-1}t} = \rho_{\alpha(t)} + O(\varepsilon)
\end{equation}
Here are some possible examples:
\begin{enumerate}
\item
\underline{Driven multilevel systems}: we consider then Markov jump processes on the possible energy levels of a system. These
can be obtained from weak coupling limits of finite quantum systems in contact with thermal reservoirs as derived for example in \cite{LebSpo}.  When they are driven, population inversions can occur such as in lasers where the protocol $\alpha$ would refer to the pumping amplitude (together with environment temperature and volume of the cavity).  The stationary level statistics $\rho_\alpha$ is then no longer given by the Boltzmann equilibrium distribution; rather $\rho_\alpha$ depends non-trivially on $\alpha$.  See also \cite{jir} for calorimetric considerations.

\item \underline{Driven Brownian particles}: that is the set-up for colloids on which rotational forces act in a viscous equilibrium medium. The dynamical variable is the position of the colloid. We will see the formalism in the next section; the protocol $\alpha(t)$ regulates conservative and non-conservative fields, together with the temperature of the medium.

\item \underline{Diffusion processes under hydrodynamic scaling}:  these appear as infinite dimensional versions of the previous example.  They can represent the rescaled dynamics of interacting particle systems at least when the hydrodynamic scaling is diffusive, \cite{lar,roma05}.  They work mostly under the assumption of local equilibrium and the protocol $\alpha(t)$ regulates the temperature and the external conservative fields.  We will see an example in Section \ref{mab}, but the set-up is more generally that of standard irreversible thermodynamics.
\end{enumerate}

\subsection{Within irreversible thermodynamics}
There is, as far as we know, not a unique notion of nonequilibrium entropy.  Within the set-up of irreversible thermodynamics however the entropy of a macroscopic system remains defined in local equilibrium.  The system can be driven either in the bulk or from the boundary but it is assumed that the ideas of heterogeneous equilibrium remain in place.
In particular that means that there is well-defined entropy density
$s(r,t)$, $r\in \Omega$ of the system in volume $\Omega$ satisfying a balance equation
\begin{equation}\label{irrt}
  \frac{\partial s}{\partial t} + \mbox{ div }  J_s(t) = \sigma(t)
\end{equation}
in terms of the entropy flux $J_s(t)$ to the environment and the entropy production $\sigma(t)$.  We refer to \cite{deGM} for the general theory, phenomenology and concrete realizations.  Later sections will also add details within specific set-ups but here we only sketch the line of the argument.\\
The time-dependence in \eqref{irrt} arises from two sources.  First from the thermodynamic transformation $\alpha(t)$ as we had above, and secondly from the time-evolution of the macroscopic variables
$\nu(r,t)$
which itself depends again on the protocol $\alpha(t)$.  These dynamical variables are mostly defined in terms of density, velocity and/or energy profiles through the volume of the system so that the time-evolution is obtained from a continuity equation together with constitutive equations.  That also implies definitions of fluxes $J_\alpha^i(r,t)$ and thermodynamic forces $X_\alpha^i(r,t)$ which define the (local) entropy production (rate)
\[
\sigma(t) = \sigma_{\alpha(t)}(r,t) = \sum_i J_\alpha^i(r,t)\, X_\alpha^i(r,t) \geq 0
\]
as a sum over the various channels $i$ of dissipation.  Similarly, the entropy flux $J_s$ in \eqref{irrt} is defined in terms of the energy and the particles  crossing the boundary $\partial \Omega$ taking into account also the boundary conditions that can be part of the control $\alpha$.\\

The argument starts by considering a fixed choice $\alpha$ of the control (not depending on time).  Then it makes sense to ask for the stationary profile $\nu_\alpha(r)$, time-invariant solution of the local equilibrium hydrodynamics. When evaluating \eqref{irrt} for that profile we have
\[
\mbox{div  } J_s(\nu_\alpha(r)) = \sigma_\alpha(\nu_\alpha(r))
\]
meaning that the stationary entropy production is entirely given in terms of the rate of change of the entropy in the environment.    We can write that for each value $\alpha(t)$ along the protocol and hence we can add their difference to \eqref{irrt} without change:
\begin{equation}\label{irrts}
  \frac{\id s}{\id t} =  \mbox{ div } J_s(\nu_{\alpha(t)}(r)) -\mbox{ div }  J_s(t) + \sigma(t) - \sigma_{\alpha(t)}(\nu_{\alpha(t)}(r))
\end{equation}
One could expect that under quasi-static transformations \eqref{qsc},
$\sigma(t) - \sigma_{\alpha(t)}(\nu_{\alpha(t)}(r))$ yields a negligible contribution, yet this is not the case in general. Although
$\sigma(t) - \sigma_{\alpha(t)}(\nu_{\alpha(t)}(r)) = O(\ve)$, \emph{cf.}~\eqref{qsc}, its total contribution to the thermodynamic process is \[
  \int_0^{\tau / \ve} \bigl[
  \sigma(t) - \sigma_{\alpha(t)}(\nu_{\alpha(t)}(r)) \bigr]\,\id t
  = O(1)
\]
and hence it remains relevant. To obtain both the Clausius equality in the quasi-static limit and the inequality beyond that limit, we would rather need the relation
$\sigma(t) - \sigma_{\alpha(t)}(\nu_{\alpha(t)}(r)) = O(\ve^2) \geq 0$, which essentially amounts to the validity of a minimum entropy production principle. However, it is well known that such an (approximative) variational principle is restricted to the linear irreversible regime in which the fluxes
$J_\al^i(r,t)$ are linearly dependent on the forces $X_\al^i(r,t)$, see~\cite{deGM}.
In that regime the sum of the last two terms in~\eqref{irrts} after time-integration remains negligible and we can write the quasi-static entropy balance equation
\[
\frac{\partial s}{\partial t}  +  \text{ div } [J_s(t) - J_s(\nu_{\alpha(t)})] = 0
\]
which is a Clausius equality for the entropy of the system under quasi-static evolutions in standard irreversible thermodynamics.  The entropy flux got renormalized by subtracting the instantaneous stationary entropy flux.
Note however, that even in the linear irreversible framework the above equality does not immediately extend to an inequality for non quasi-static processes, the basic reason being that a strongly non quasi-static evolution easily breaks the restrictions of the linear irreversible framework.
A precise argument valid also outside the set-up of irreversible thermodynamics but for close-to-quasi-static evolutions follows around eq. \eqref{17}.\\
The new idea of the paper is to formulate a modification of the minimum entropy production principle which does remain valid arbitrarily far from equilibrium, at least for diffusions without inertial degrees of freedom. Since it naturally follows from dynamical fluctuations of driven Brownian particles, we first formulate our proposal within the ``single-particle'' framework where the entropy function simply boils down to the Shannon entropy (Section~\ref{modc}). That however enables a direct generalization to a class of driven macroscopic diffusions at local equilibrium, including those driven by boundary thermodynamic forces (a specific scenario will be discussed in Section~\ref{mab}).

\section{Model and previous results}\label{prev}

We consider a Brownian particle moving in some (possibly multidimensional) spatially and temporally dependent environment, which is coupled to a heat bath and driven by non-conservative forces. It is introduced by the It\^o stochastic equation
\begin{equation}
  \id x_t = \be D(f - \nabla U)\,\id t + \nabla \cdot D\,\id t
  + (2 D)^{1/2} \id W_t
\end{equation}
with $W_t$ the standard Wiener process and
for some driving force $f = f_t(x)$, potential landscape $U = U_t(x)$, inverse temperature $\be = \beta_t=  (k_B T_t)^{-1}$ and positive diffusion matrix
$D = D_t(x)$. In accordance with the Einstein relation, the  matrix $\be D$ corresponds to the particle's mobility.
The time-dependent probability distribution evolves according to the Smoluchowski equation
\begin{equation}\label{smol}
  \frac{\partial \mu_t}{\partial t} + \nabla\cdot J_{\mu_t}^U = 0
\end{equation}
with the current density
$J_\mu^U = \mu\, \be D\,(f - \nabla U) - D\,\nabla\mu$. That is a version of the Master equation \eqref{mas} for $\alpha=(f,U,\beta,D)$. The time-dependence will often not be explicitly denoted. We have added the superscript $U$ to indicate the dependence on the potential landscape which plays an essential role in our argument below.
We put no restrictions on the space-time dependence of the model parameters
except for the usual smoothness and ergodicity conditions. In particular, we assume that the potential $U$ is sufficiently confining so that normalizable solutions to the stationary equation $\nabla\cdot J_{\mu}^U = 0$ always exist. We also assume either free boundary conditions at infinity, or that the Brownian particle moves on the one-dimensional circle (= periodic boundary conditions).  The meaning of probabilities is here most simply that $\mu_t$ represents the density of many independent particles at position $x$; that is also why the Shannon entropy $ S(\mu) = -\int \mu \log\mu\,\id x$ ($k_B \equiv 1$) is here the relevant thermodynamic entropy.  Models of interacting particles will be treated in Section \ref{mab}.\\

The main point of departure in all that follows is the balance equation: under \eqref{smol},
\begin{equation}\label{balance}
  \frac{\id S(\mu)}{\id t} = \be\,\frac{\de Q^U(\mu)}{\id t} + \si^U(\mu)
\end{equation}
with mean incoming heat flux
\begin{equation}
\begin{split}
  \frac{\de Q^U(\mu)}{\id t} &=
  \Bigl\langle (\nabla U - f) \circ \frac{\id x_t}{\id t} \Bigr\rangle_\mu
\\
  &= \int (\nabla U - f) \cdot J_\mu^U\,\id x
\end{split}
\end{equation}
and entropy production rate
\begin{equation}\label{entprod}
  \si^U(\mu) = \int J_\mu^U \cdot (\mu D)^{-1} J_\mu^U\,\id x \geq 0
\end{equation}
The balance \eqref{balance} can be further improved because, in the presence of a nonequilibrium driving even for arbitrarily slow processes,
the entropy production rate $\si^U(\mu)$ remains nonzero, indicating the irreversibility. Time-integrating \eqref{balance} over long times (as for quasi-static processes \eqref{qst}) would result in infinities. How to renormalize that balance equation while keeping physically interesting excess quantities is the first challenge in extending the Clausius heat theorem.  It leads to a number of different proposals as we now first briefly review.

\subsection{Hatano-Sasa approach}\label{hsa}

Hatano and Sasa~\cite{hs} have proposed to separate from the total heat
its ``house-keeping'' component as corresponding to the (as far as we see, artificial) force
$f - \nabla U - \be^{-1}\nabla \log\rho$ with $\rho$ the stationary density solving the stationary equation
$\nabla \cdot J_\rho^U = 0$. This house-keeping heat has the instantaneous expectation
\begin{equation}\label{house-keep}
\begin{split}
  \frac{\de Q^{\text{hk}}(\mu)}{\id t} &= \Bigl\langle
  (\nabla U - f + \be^{-1}\nabla\log\rho) \circ \frac{\id x_t}{\id t} \Bigr\rangle_\mu
\\
  &= -\int J_\rho^U \cdot (\rho\,\be D)^{-1} J_\mu^U\,\id x
\end{split}
\end{equation}
to be compared with \eqref{entprod}.
That definition of the house-keeping heat relates to time-reversal considerations, see \cite{gaw}. Observe that under the detailed balance condition which is realized for $f = 0$, the stationary density has the Gibbsian form
$\log\rho = -\be U + \text{const}$ and $J_\rho^U = 0$; hence
$\de Q^\text{hk} / \id t = 0$.
\\

The ``excess'' component $\de Q^\text{HS}$ in the decomposition
$\de Q^U = \de Q^\text{hk} + \de Q^\text{HS}$
reads
\begin{equation}\label{hs-excess}
  \frac{\de Q^{HS}(\mu)}{\id t} =
  -\be^{-1} \int \nabla\log\rho \cdot J_\mu^U\,\id x
\end{equation}
and the entropy balance~\eqref{balance} obtains the form
\begin{equation}\label{hs-balance}
  \frac{\id S(\mu)}{\id t} = \be\,\frac{\de Q^\text{HS}}{\id t}
  + \int \Bigl( J_\mu^U - \mu\, \frac{J_\rho^U}{\rho} \Bigr) \cdot
  (\mu D)^{-1} \Bigl( J_\mu^U - \mu\, \frac{J_\rho^U}{\rho} \Bigr)\,\id x
\end{equation}
Since the integral on the right-hand side is manifestly nonnegative, we arrive at the exact Clausius-type inequality
$\id S(\mu) \geq \be\, \de Q^{\text{HS}}$.

Now we take a generic thermodynamic process as in Section \ref{ttr} specified by some protocol
$\al(t) = (f_{t}, U_{t}, \be_{t}, D_t)$, $0 \leq t \leq \tau$. We consider its re-scaled variant
$\al^{(\varepsilon)}(t) = \al(\varepsilon t)$, $0 \leq t \leq \tau/\varepsilon$ in the quasi-static limit $\varepsilon \downarrow 0$ as in \eqref{qst}. Under standard conditions for the adiabatic theorem to hold true, see \eqref{qsc}, the instantaneous densities solving the Smoluchowski equation \eqref{smol} under the protocol $\al^{(\varepsilon)}(t)$ are of the form
$\mu_t^{(\varepsilon)} = \rho^{(\varepsilon)}_t + O(\varepsilon)$
where $\rho^{(\varepsilon)}_t$ is the stationary density at fixed condition
$\alpha = \al^{(\varepsilon)}(t)$.
As a consequence,
the integral on the right-hand side of~\eqref{hs-balance} is
$O(\varepsilon^2)$ and the time integral of that balance equation from
$t = t_\text{ini} = 0$ till $t = t_\text{fin} = \tau / \varepsilon$ reads
\begin{equation}\label{hs-equality}
%  S(\rho_\text{fin}) - S(\rho_\text{ini}) =
  S(\mu^{(\varepsilon)}_{\tau / \varepsilon}) - S(\mu_0) =
  \int_0^{\tau / \varepsilon} \be^{(\varepsilon)}_t \de Q^{\text{HS},(\varepsilon)}(\mu_t)
  + O(\varepsilon)
\end{equation}
yielding an equality in the limit $\varepsilon \downarrow 0$, with
$S(\mu_0) = S(\rho_\text{ini})$ and
$S(\mu^{(\ve)}_{\tau / \varepsilon}) \to S(\rho_\text{fin})$.
(Note that for a rigorous treatment some further regularity assumptions are needed to ensure time uniformity of the $\varepsilon-$expansions.)\\

As obvious from the construction, the essence of the method lies in formally replacing the actual mechanical force $f - \nabla U$ with the ``thermodynamic'' force derived from the stationary potential $-\be^{-1}\log\rho$; the heat produced by this fictitious force matches the fluctuating excess heat. However, it is not obvious how to operationally access either the house-keeping heat~\eqref{house-keep} or directly the excess heat~\eqref{hs-excess}, due to the nontrivial dependence on both densities $\rho$ and $\mu$. A ``non-calorimetric'' experimental verification has been reported in~\cite{bust04}.

A precise formulation of the Hatano-Sasa approach aided and connected with fluctuation symmetries can be found in \cite{gaw}, in particular relevant via their Example 12 on page 492 of that paper. In fact, the Hatano-Sasa work is truly a formal nonequilibrium generalization of the Jarzynski equality.  Here we have emphasized the part that fits the subject of the present paper.

\subsection{Komatsu-Nakagawa-Sasa-Tasaki approach}\label{sec:knst}

A somewhat different road has been proposed in~\cite{knst} where the heat is decomposed in such a way that its steady-state component allows for a more direct interpretation and experimental access. Next we review this method in a slightly modified way which better reveals how it is related to our approach that will be discussed later.

Here the decomposition is done by subsequently removing from the heat the integrated \emph{steady} heat flux
$\de Q^U(\rho) / \id t = \int (\nabla U - f) \cdot J_\rho^U\,\id x$, again with  $\rho$ being the stationary density,
$\nabla \cdot J_\rho^U = 0$.
By stationarity,
$-\de Q^U(\rho) / \id t = \si^U(\rho)$ and the balance relation~\eqref{balance} can be written in terms of the remaining (``excess'') heat
$\de Q^\text{ex}(\mu) = \de Q^U(\mu) - \de Q^U(\rho)$ as
\begin{equation}\label{renold}
  \frac{\id S(\mu)}{\id t} = \be\,\frac{\de Q^\text{ex}(\mu)}{\id t} +
  \si^U(\mu) - \si^U(\rho)
\end{equation}
In contrast to the previous approach, the removed steady-state flux and hence also the remaining excess heat are \emph{in principle} directly accessible. However, the difference
$\si^U(\mu) - \si^U(\rho)$ can in general take both positive and negative values unless further restrictions are imposed as explained next.

Close to equilibrium, \emph{i.e.}\ for weak non-conservative forces and small deviations from stationarity, the entropy production
$\si^U(\mu)$ as a function of density $\mu$ attains its minimum at
$\mu^* \approx \rho$ --- the  Minimum Entropy Production Principle (MEPP). Then, $\si^U(\mu) - \si^U(\rho)\geq 0$ in \eqref{renold} which gives the Clausius inequality.  More precisely, we write
$f = \la f_1$ and $\log(\mu/\rho^{(\la)}) = \ze g_1$ with $\la$ and $\ze$ control parameters, and we expand the entropy production difference around
$\ze = \la = 0$. For any (smooth) $f_1$ and $g_1 \neq 0$,
\begin{equation}\label{17}
   \si^U(\mu) - \si^U(\rho^{(\la)}) = \ze^2 B(\la) + O(\ze \la^2, \ze^3)
\end{equation}
with some function $B(\la) \geq b > 0$ in a neighborhood of $\la = 0$. In particular, if taking $\la = O(\ze)$, we obtain
$\si^U(\mu) > \si^U(\rho^{(\la)})$ in a neighborhood of $\ze = 0$. Whenever this variational principle can (at least approximately) be verified, the balance equation~\eqref{renold} yields the generalized Clausius-type inequality
$\id S(\mu) \geq \be\,\de Q^\text{ex}(\mu)$.

Finally we combine the close-to-equilibrium expansion with the quasi-static expansion \eqref{qst}--\eqref{qsc}. For a thermodynamic process defined by a re-scaled protocol
$\al^{\varepsilon}(t)$, $0 \leq t \leq \tau / \varepsilon$ as before, we have
$\ze = O(\varepsilon)$ by the adiabatic theorem, and therefore the entropy balance equation obtains the time-integrated form~\cite{knst}
\begin{equation}
  S(\mu^{(\varepsilon)}_{\tau / \varepsilon}) - S(\mu_0) =
  \int_0^{\tau / \varepsilon} \be_t\,\de Q^{\text{ex},\varepsilon}(\mu_t) +
  \varepsilon C(\la) + O(\la^2,\varepsilon^2)
\end{equation}
with some $C(\la) \geq c > 0$ in a neighborhood of $\la = 0$. The second term on the right-hand side vanishes in the quasi-static limit
$\varepsilon \downarrow 0$, whereas the remaining (indefinite) corrections only appear in quadratic order around equilibrium. An exact inequality is again verified, e.g., for $\la = O(\varepsilon)$ in a neighborhood of $\varepsilon = 0$, with the strict equality in the (quasi-static and equilibrium) limit
$\varepsilon \downarrow 0$.

Although this approach deals with physically more directly accessible quantities, its drawback obviously lies in its limitation to the close-to-equilibrium regime, and we have seen how it originates from the violation of the MEPP. But a deeper understanding of that minimum entropy production principle arises from the study of dynamical fluctuations, \cite{min}.  The next section and main result of the paper is in fact directly inspired by looking at the rate function of occupation time statistics.

\section{Modified Clausius theorem}\label{modc}

Our strategy to overcome the above restrictions will be to first find a modification of the minimum entropy production principle which remains \emph{exactly} valid far from equilibrium, and then to apply the corresponding decomposition scheme for heat.
We show that this problem has a surprisingly simple solution, at least within the framework of overdamped diffusions to which we restrict here.
This approach is motivated and is intimately related to the theory of dynamical fluctuations; this connection will be explained afterwards.

\subsection{Heat decomposition from an exact MEPP}
\label{sec:modified}

Given the entropy production $\si^U(\mu)$ by equation~\eqref{entprod}, we now consider it as a functional of the potential landscape $U$, and with the density $\mu$ and all other system parameters fixed. It is easy to check that this functional is convex and
\begin{equation}\label{canonical}
  \frac{\de \si^U(\mu)}{\de U} =
  2\be\,\nabla \cdot J^U_\mu
\end{equation}
Hence its minimizer $V = V^\mu$ satisfies $\nabla \cdot J^V_\mu = 0$, \emph{i.e.}, it is such a potential landscape that makes the density $\mu$ stationary.
The solvability and uniqueness of this \emph{inverse} stationary problem has been studied in the context of Donsker-Varadhan theory of large deviations~\cite{dds}. The resulting inequality
$\si^U(\mu) \geq \si^{V}(\mu)$
can be interpreted as a modified MEPP, stating that:
\emph{Among all possible potential landscapes, the system at a given state dissipates the least at stationary conditions.}

This ``complementary'' variant of the usual (approximative) MEPP principle provides a variational formulation of the inverse stationary problem, \emph{i.e.}, to find a potential $U$ that makes a given density $\mu$ stationary. In contrast to the original principle (or hypothesis), our version remains \emph{exactly} valid arbitrarily far from equilibrium. We remark that the optimal potential $V^\mu$ is a non-local functional of the density $\mu$, in the sense of long-range dependencies in the linear response function
$\chi(x,y) = \partial V^\mu(x) / \partial\mu(y)$ \cite{MNS09}.\\

We now propose to decompose the heat via the integrated steady fluxes obtained from the inverse (instead of the direct as before) stationary problem. The new heat decomposition reads
\begin{equation}\label{mex}
  \de Q^U(\mu) = \de Q^V(\mu) + \de Q^\text{mex}(\mu)
\end{equation}
where
\begin{equation}
  \frac{\de Q^V(\mu)}{\id t} =
  \int (\nabla V - f) \cdot J_\mu^V\,\id x =
  -\si^V(\mu)
\end{equation}
is a modified steady-state heat flux but now with respect to the steady state specified by the density $\mu$ (and under the potential $V = V^\mu$ accommodated so that it provides a stationary condition for that $\mu$).
The modified excess component is
\begin{equation}\label{mex-explicit}
  \frac{\de Q^\text{mex}(\mu)}{\id t} =
  \int (\nabla U - f) \cdot (J_\mu^U - J_\mu^{V})\,\id x
\end{equation}
and the entropy balance equation~\eqref{balance} obtains the form
\begin{equation}\label{balance-new}
  \frac{\id S(\mu)}{\id t} = \be\, \frac{\de Q^\text{mex}(\mu)}{\id t} +
  \si^U(\mu) - \si^V(\mu)
\end{equation}
which is very similar to the relation~\eqref{renold}.
However, in contrast to the latter, the modified MEPP now implies the exact Clausius-type inequality $\id S(\mu) \geq \be\, \de Q^\text{mex}(\mu)$ without any close-to-equilibrium restrictions. Upon time-integration along some thermodynamic protocol $\al(t)$, $0 \leq t \leq \tau$, we get
\begin{equation}\label{mcm}
  S(\mu_\tau) - S(\mu_0) \geq \int_0^\tau \be_t\,\de Q^\text{mex}(\mu_t)
\end{equation}

The quasi-static equality follows by adjusting the argument from the previous sections. Again with $\varepsilon$ the quasi-static control parameter we write the time-integral of the balance equation~\eqref{balance} in the form
\begin{equation}\label{balance-ren}
  S(\mu^{(\varepsilon)}_{\tau / \varepsilon}) - S(\mu_0) =
  \int_0^{\tau / \varepsilon} \be^{(\ve)}_{t}\,\de Q^{\text{mex},\ep}(\mu_t^{(\varepsilon)})
  + \int_0^{\tau / \varepsilon} (\si^{{U_t},\varepsilon} - \si^{{V_t},\varepsilon})(\mu^{(\varepsilon)}_t)\,\id t
\end{equation}
where $\mu_t^{(\varepsilon)}$ is the solution of the Smoluchowski equation \eqref{smol} with the system parameters according to the protocol $\al^{(\varepsilon)}(t) = \al(\varepsilon t)$.  The dynamical dependence of the entropy production $\si$ on $\ve$ is via $\al^{(\ve)}(t)$, i.e., it depends on the conditions at time $t$.  There is however also the dependence on the potential, which enables introducing an $\ve-$expansion for the difference in the second integral, as now follows.  We assume that this expansion can be made uniformly with respect to the non-rescaled time  $t \in [0,\tau]$.\\
Using that
$\mu^{(\varepsilon)}_t = \rho^{(\varepsilon)}_t + O(\varepsilon)$ we apply the implicit function theorem to the inverse stationary problem to get
$V_t^{(\varepsilon)} = U_t^{(\varepsilon)} + O(\varepsilon)$. Hence
$\si^{U_t,\varepsilon}(\mu_t^{(\varepsilon)}) = \si^{V_t,\varepsilon}(\mu_t^{(\varepsilon)}) + O(\varepsilon^2)$ since, by construction,
the functional $\si^{U,(\varepsilon)}(\mu_t^\varepsilon)$ takes its minimum at $U = V_t^{(\varepsilon)}$.
Therefore the second time-integral in~\eqref{balance-ren} is $O(\varepsilon)$ and we get in the quasi-static limit
\begin{equation}\label{mcmq}
  S(\rho_\text{fin}) - S(\rho_\text{ini}) =
  \lim_{\varepsilon \downarrow 0} \int_0^{\tau / \varepsilon} \be^{(\varepsilon)}_t\,\de Q^{\text{mex},\varepsilon}(\mu_t^{(\varepsilon)})
\end{equation}
This variant \eqref{mcm} and \eqref{mcmq} of the nonequilibrium Clausius heat theorem is our first (main) result.

\subsection{Connection to dynamical fluctuations}\label{dync}

It is by now rather well understood how various equilibrium and nonequilibrium variational principles originate from the study of (either static or dynamic) fluctuations, examples being provided by Einstein's static and Onsager-Machlup's dynamic fluctuation theories which have obtained many generalizations, \emph{e.g.}, in the context of fluctuating hydrodynamics, \cite{roma05}. Here we make use of results obtained in~\cite{dds} to relate our renormalized entropy production to a fluctuation potential so that the origin of the modified MEPP which has been a key ingredient in our argument, gets further clarified.

We study the question of how plausible it would be to observe the intermediate density $\mu_t$ the system exhibits at each time $t$, as the actual empirical occupation statistics with respect to the \emph{steady} dynamics at ``frozen'' parameters
$f_t, U_t, \be_t, D_t$.
From the observed realization $(x_{t+s})_{s=0}^{\tau}$ of that steady process we extract the time-averaged occupation density
\begin{equation}
  p_t^\tau(z) = \frac{1}{\tau} \int_0^{\tau} \de(x_{t+s} - z)\,\id s
\end{equation}
By the assumed ergodicity, $p_t^\tau(z)$ \emph{typically} converges to $\rho_t$ for
$\tau \to +\infty$. The general result about the probability of \emph{untypical} empirical densities reads that it has the exponential asymptotic form ($\tau \to +\infty$),
\begin{equation}\label{donvar}
  \text{P}(p_t^\tau \simeq \mu_t) = \exp\,[-\tau I^{U_t}(\mu_t) + o(\tau)]
\end{equation}
For the diffusion processes under consideration the rate function has the form, see~\cite{dds},
\begin{equation}\label{ld}
  I^{U}(\mu) = \frac{1}{4}\,[\si^{U}(\mu) - \si^{V}(\mu)]\,,\qquad
  \nabla \cdot J_\mu^V = 0
\end{equation}
By construction, $I^U(\mu) \geq 0$ with equality if and only if
$\nabla \cdot J_\mu^U = 0$. Fixing $U$ and taking density $\mu$ as a variable, that provides the approximative MEPP principle invoked in Section~\ref{sec:knst} since, in the notation used therein,
$\si^V(\mu) = \si^U(\rho) + O(\ep \la^2, \ep^3)$,
\cite{min}. On the other hand, fixing $\mu$ and varying $U$, we immediately obtain the MEPP introduced in Section~\ref{sec:modified} and applied in our new heat decomposition scheme.

Combining eqs.~\eqref{balance-new} and~\eqref{ld}, the modified excess heat can be written in the form
\begin{equation}
  \be\,\frac{\de Q^\text{mex}(\mu)}{\id t} =
  \frac{\id S(\mu)}{\id t} - 4 I^U(\mu)
\end{equation}
which directly links the probability of rare dynamical events to the violation of the extended Clausius theorem.
That summarizes the major point of the paper.

\section{Macroscopic boundary-driven diffusion}\label{mab}

So far we have restricted ourselves to finite diffusions which are most easily interpreted for independent particles.  Yet our formalism allows a rather straightforward extension to a class of macroscopic bulk- and boundary- driven diffusion systems. The reason is that the macroscopic density profile evolves with time according to a non-linear variant of the Smoluchowski equation, with a density-dependent diffusion matrix derived from microscopic details of particle interactions. Whereas this extra nonlinearity does not require any essential changes in the construction of the modified excess heat, the incorporation of (possibly time-dependent) boundary conditions demands some modifications that will be explained below.\\

As a specific model we consider a Brownian fluid described by the density profile
$n = n_t(r)$ in a bounded domain $\Om$. We assume the fluid to be
\emph{locally} always at equilibrium with a heat bath at inverse temperature
$T_t = 1 / (k_B \be_t)$, but \emph{globally} being driven out of equilibrium via maintaining a spatially inhomogeneous chemical potential profile at the system's boundary. (Adding a bulk field would also be possible.)  The (thermo)dynamics of the system is specified via (1) a free energy density functional $\phi(n_t(r),T_t)$ describing the local equilibrium in the bulk, via (2) a positive-definite mobility matrix $\Lambda(n_t(r))$, and via (3) a boundary profile of the chemical potential
$\bar\al_t(r)|_{r \in \partial\Om}$. Adding the variable control field
$U_t(r)$ vanishing on the boundary,
$U_t(r) = 0$ for all $r \in\partial\Om$, the density profile evolves following the continuity equation
\begin{equation}\label{density-macro}
  \frac{\partial n_t(r)}{\partial t} + \nabla \cdot J_t^{U}(r) = 0
\end{equation}
with the current density functional
\begin{equation}\label{current-macro}
  J_t^U(r) = -\La(n_t(r))
  \nabla\Bigl(\frac{\partial\phi(n_t(r),T_t)}{\partial n} + U_t(r) \Bigr)
\end{equation}
and the boundary condition
\begin{equation}
  \frac{\partial \phi(n_t(r),T_t)}{\partial n}\Bigr|_{r \in \partial\Om} =
  \bar\al_t(r)|_{r \in \partial\Om}
\end{equation}
For simplicity, we restrict ourselves (as indicated above) to thermodynamic processes corresponding to time-dependent bulk temperature $T_t$, the boundary chemical potentials $\bar\al_t(r)$ and the control field $U_t(r)$, though various generalizations are easy to obtain.\\

For the total free energy
$\Phi_t(r) = \int_\Om \phi(n_t(r),T_t)\,\id r$ we have
\begin{equation}\label{free-energy}
  \frac{\id\Phi_t}{\id t} =
  - S_t\,\frac{\id T_t}{\id t} +
  \int_\Om \nabla\al_t(r) \cdot J^U_t(r)\,\id r
  - \oint_{\partial\Om} \bar\al_t(r)\,J^U_t(r) \cdot \id\Sigma
\end{equation}
where $S_t = -\int_\Om \partial\phi(n_t(r),T_t) / \partial T\,\id r$ is the total entropy and
$\al_t(r) = \partial\phi(n_t(r),T_t) / \partial n$ is the local chemical potential. The last term is over the boundary of $\Omega$ and arises, together with the second term, from partial integration.\\
To fit the equation~\eqref{free-energy} in the standard framework of irreversible thermodynamics, we rewrite it in terms of the total energy
$\caE^U_t = \Phi_t + T_t S_t + \int U_t(r)\, n_t(r)\,\id r$  (including the energy of the control field), as
\begin{equation}\label{ebe}
\begin{split}
  T_t\,\frac{\id S_t}{\id t} &=
  \frac{\id\caE^U_t}{\id t} - \int_\Om \frac{\id U_t(r)}{\id t}\,n_t(r)\,\id r
  + \int_\Om \nabla(\al_t + U_t)(r) \cdot J^U_t(r)\,\id r
  - \oint_{\partial\Om} \bar\al_t(r) \cdot J_t^U(r)\,\id\Sigma
\end{split}
\end{equation}
where the first integral on the right-hand side is (minus) the work of the control field and the last integral is the convective energy outflow. Hence, there has appeared the total (incoming) heat
\begin{equation}
  \frac{\de Q^U_t}{\id t} =
  \frac{\id\caE^U_t}{\id t} - \int_\Om \frac{\id U_t(r)}{\id t}\,n_t(r)\,\id r
  - \oint_{\partial\Om} \bar\al_t(r) \cdot J_t^U(r)\,\id\Sigma
\end{equation}
from both the bulk and the boundary reservoirs. Similarly, we recognize in \eqref{ebe} also the entropy production functional
\begin{equation}
\begin{split}
  \si^U_t &= -\frac{1}{T_t} \int_\Om \nabla(\al_t + U_t)(r) \cdot J_t^U(r)\,\id r
\\
  &= \frac{1}{T_t} \int_\Om J^U_t(r) \cdot \La^{-1}(n_t(r)) J^U_t(r)\,\id r \geq 0
\end{split}
\end{equation}
so that we finally get the standard entropy balance
\begin{equation}
  \frac{\id S_t}{\id t} = \frac{1}{T_t} \frac{\delta Q^U_t}{\id t} + \si^U_t
\end{equation}
We have arrived at the diffusion scheme of \eqref{balance}--\eqref{entprod}.\\

Now we are ready to apply the analogous heat decomposition scheme as in Section~\ref{sec:modified}. Since
$\si^U$ is a convex quadratic functional of the control field $U$ and using that
\begin{equation}
  \frac{\de\si^U}{\de U(r)} = \frac{2}{T_t} \nabla \cdot J^U(r)
\end{equation}
everywhere in the interior of volume $\Om$ (recall that by definition $U$ vanishes on the boundary $\partial\Om$)
we again obtain the modified MEPP, according to which the stationarity condition
$\nabla \cdot J^U = 0$ for the field $U$ is equivalent to the minimization of the entropy production functional $\si^U$ on the space of all (smooth) fields
$U$, $U(r)|_{r \in \partial V} = 0$. Skipping the rigorous treatment of this variational problem, we now continue exactly as in Section~\ref{sec:modified} to define the modified excess heat, by removing from the heat its steady flux corresponding to the reference stationary dynamics under the control field
$V_t$ for which $\nabla \cdot J_t^{V_t} = 0$. Hence, with $U_t$ the imposed field and $V_t$ its stationary modification,
\begin{equation}
  \de Q_t^\text{mex} = \de Q_t^{U} - \de Q_t^{V}
\end{equation}
satisfies the generalized Clausius inequality
\begin{equation}
  S_\tau - S_0 \geq \int_0^\tau \de Q_t^\text{mex}
\end{equation}
which again becomes an equality in the quasi-static limit, by the same arguments as before.
The result is similar but different from the one in \cite{roma12,roma13} by our use of the modified minimum entropy production principle and the treating here of a somewhat broader class of free energy functionals and possible driving mechanisms.
As we have shown the reason is that the arguments of Section \ref{modc} allow for an almost immediate extension to macroscopic bulk-driven diffusions of the form~\eqref{smol} in which the diffusion matrix $D$ becomes density-dependent.

\section{Time-reversal considerations}\label{trco}

We have seen that the proper definition of the (finite) heat component truly associated with the \emph{transformation} between steady states rather than with maintaining them, is the key ingredient in virtually all generalizations of the Clausius heat theorem to nonequilibrium. Nevertheless, other types of formal extensions may of course be invented. The present section gives such other point of view, using the different symmetry properties of both heat components under a suitable reversal operation, staying on the one hand close to the Hatano-Sasa approach of Section \ref{hsa}, as also formulated in \cite{gaw}, and on the other hand applying a time-reversal operation intimately related to our construction of the modified excess in Section \ref{modc}.\\

In the original formulation, our model is specified by the protocol
$(U_t; f_t, \be_t, D_t)_{t=0}^\tau$ of externally controlled parameters, and the trajectory
$(\mu_t)_{t=0}^\tau$ of evolved distributions obeys the Smoluchowski equation \eqref{smol}. We now take a {\it dual} point of view and we see the latter,
$\dot\mu + \nabla \cdot J_\mu^U = 0$,\ as an equation for the time-dependent potential $(U_t)_{t=0}^\tau$ such that a prescribed trajectory of densities
$(\mu_t)_{t=0}^\tau$ actually takes place. Obviously, this is a non-stationary generalization of the inverse problem considered in Section~\ref{sec:modified}.
According to~\eqref{canonical}, this can be seen as a Legendre transformation from the potential $U$ to its conjugated field
$\de\si^U(\mu) / \de\mu$ as generated by the functional $\si^U(\mu)$. This inverse problem was introduced in~\cite{MNW08a} in the context of (non-stationary) dynamical fluctuations.

In our new picture, the time-evolved system is fully described in terms of the ``imposed'' trajectory $(\mu_t; f_t,\be_t,D_t)_{t=0}^\tau$ of both the states and of the protocol, whereas the potential $(U_t)_{t=0}^\tau$ becomes a dependent quantity determined from the Smoluchowski equation (up to a constant). We now consider  reversed trajectories (for simplicity, let $D_t = 1$ in the sequel)
\begin{equation}\label{reversal}
  (\mu_t;\, f_t,\beta_t)^\dagger = (\mu_{\tau-t};\, f_{\tau-t}^\dagger,\beta_{\tau-t})\,,\qquad
  f_t^\dagger = - f_{\tau-t} + \nabla \psi_{\tau - t}
\end{equation}
where $\psi_t$ can be chosen arbitrarily. Then we check that if $U_t$ is the solution to the (inverse-problem) Smoluchowski equation for
$(\mu_t; f_t,\be_t)$ then
\begin{equation}\label{potential-rev}
  U_t^\dagger = \psi_{\tau - t} - U_{\tau - t} +
  2 \be_{\tau - t}^{-1} \log \mu_{\tau - t}
\end{equation}
is the solution to the  Smoluchowski equation for $(\mu_t; f_t,\be_t)^\dagger$. Indeed,
from~\eqref{potential-rev} we have
\begin{equation}\label{current-rev}
\begin{split}
  J_t^\dagger \equiv J_{\mu_t^\dagger}^{U_t^\dagger} &=
  \mu_t^\dagger \be_t^\dagger \bigl( f_t^\dagger - \nabla U_t^\dagger \bigl) -
  \nabla \mu_t^\dagger
\\
  &= -\mu_{\tau - t} \be_{\tau - t} \bigl( f_{\tau - t} - \nabla U_{\tau - t}
  \bigr) + \nabla \mu_{\tau - t}
\\
  &= -J_{\tau - t}
\end{split}
\end{equation}
and since $\dot\mu^\dagger_t = -\dot\mu_{\tau - t}$, the equality
$\dot\mu^\dagger_t + \nabla \cdot J_t^\dagger = 0$ is proven.
As a consequence, the heat in the time-reversed scenario~\eqref{reversal} is
\begin{equation}
\begin{split}
  \frac{\delta Q_t^\dagger}{\id t} &=
  \int (\nabla U_t^\dagger - f_t^\dagger)\cdot J_t^\dagger\,\id x
\\
  &= \frac{\delta Q_{\tau-t}}{\id t} +
  2 \beta_{\tau-t}^{-1}
  \int \nabla \log \mu_{\tau-t} \cdot J_{\tau-t}\,\id x
\end{split}
\end{equation}
and by time integration we obtain our final result
\begin{equation}
\begin{split}
  S(\mu_\tau) - S(\mu_0) &=
  -\int_0^\tau \id t \int \nabla\log\mu_t \cdot J_t\,\id x
\\
  &= \frac{1}{2} \int_0^\tau
  (\be_t\,\de Q_t - \be^\dagger_t\,\de Q^\dagger_t)
\end{split}
\end{equation}
true without further conditions.\\
That extended form of Clausius \emph{equality} staying valid beyond the quasi-static regime is our second main result. In words, it relates the change of a system's (Shannon) entropy to the time-reversal antisymmetric component of the entropy flux. There is no need to explicitly mention an excess heat since that is now replaced by the antisymmetrization procedure. Note however that we have employed here a non-standard time-reversal operation~\eqref{reversal} which ensures the exact current reversal~\eqref{current-rev}. The inherent arbitrariness (via the free ``gauge potential'' $\psi$) is related to the physical non-uniqueness of the decomposition of the force into gradient and non-gradient components.

\section{Conclusion}

We have derived an extended Clausius heat theorem for overdamped diffusion processes, which remains valid arbitrarily far from equilibrium. The idea has been to modify previously adopted heat decomposition schemes according to an exact modification of the Minimum Entropy Production Principle, which naturally emerges from the structure of dynamical fluctuations. Our removed ``house-keeping'' heat does correspond to an operationally accessible steady-state flux, though with respect to a steady state obtained from solving a modified (inverse) stationary problem. Moreover, the generalized Clausius (in)equality proposed here is not restricted to a close-to-equilibrium regime. We have also discussed an extension to a class of driven macroscopic diffusions.  In fact, the structure of the argument remains exactly intact for diffusing fields where the fields correspond to  macroscopic profiles of interacting particles in a diffusive continuum approximation.

The basic reason for the restriction to overdamped diffusions (finite or infinite dimensional) is due to the fact that the fluctuation relation~\eqref{donvar} in terms of the entropy production is no longer valid beyond the diffusion regime. For discrete jump process like, \emph{e.g.}, chemical networks, the rate function for occupation times does retain a similar structure as in~\eqref{donvar} but with the entropy production $\si^U(\mu)$ being replaced with another function called dynamical activity (or ``traffic''), which becomes essentially different from the entropy production unless the system is close to equilibrium and the deviations from stationarity are small~\cite{min,MNW08a}. Within the proposed scheme, we still can check that the Clausius equality remains approximately valid up to first order around detailed balance, which (partially) reproduces the results from~\cite{knst}.
It still remains to be seen how the dynamical fluctuation framework proposed above can be used to analyze heat processes for strongly nonequilibrium jump processes and in systems with inertial degrees of freedom.

\begin{acknowledgments}
This work started from discussions with Keiji Saito, Hal Tasaki, Shin-Ichi Sasa and Naoko Nakagawa in January--February 2012.
CM wishes to thank the hospitality of the Yukawa institute in Kyoto and of Hisao Hayakawa in particular. KN thanks Shin-Ichi Sasa for fruitful discussions and acknowledges the support from the Grant Agency of the Czech Republic, Grant no.~P204/12/0897.\\
The results of the present paper were first reported on 6 June 2012 at the Isola del Giglio during the workshop on
{\it Non-equilibrium fluctuation-response relations}, June 5--8, 2012, in parallel with the talk by G.~Jona-Lasinio on the work in \cite{roma12,roma13}.  In that same context we are also grateful for private communication with K. Gawedzki.

\end{acknowledgments}

% -----------------------------------------------

\end{document}